\documentclass[12pt]{article}
\usepackage{a41}
\usepackage{floatfig,epsfig}
\usepackage{rotating}
\newcounter{my}

\newcommand{\la}[1]{\label{#1}}
\newcommand{\re}[1]{\ (\ref{#1})}
\newcommand{\nn}{\nonumber}
\newcommand{\ed}{\end{document}}
\newcommand{\be}{\begin{equation}}
\newcommand{\ee}{\end{equation}}

\newcommand{\ba}{\begin{eqnarray}}
\newcommand{\ea}{\end{eqnarray}}
\newcommand{\baz}{\begin{eqnarray*}}
\newcommand{\eaz}{\end{eqnarray*}}
\newcommand{\bb}{\begin{thebibliography}}
\newcommand{\eb}{\end{thebibliography}}
\newcommand{\ct}[1]{${\cite{#1}}$}

\newcommand{\bi}[1]{\bibitem{#1}}

\begin{document}
\initfloatingfigs
\sloppy
\thispagestyle{empty}

\vspace{1cm}

\mbox{}

\vspace{2cm}

\begin{center}
{\Large {\bf  Instantons and Polarized\\ [2mm]
Structure Functions}}
\footnote{ The  talk presented at the
Workshop "Deep Inelastic Scattering off Polarized Targets:
Theory Meets Experiment",
September 1997, DESY-Zeuthen}\\[2cm]
{\large N.I.Kochelev}\\[0.2cm]
{\it
 Bogoliubov Laboratory of Theoretical Physics,\\
Joint Institute for Nuclear Research,\\
RU-141980 Dubna, Moscow region, Russia}\\ [0.3cm]

\end{center}
\vspace{2cm}


\begin{abstract}
The contribution of  the  quark-quark
and quark-gluon interaction induced by instantons  to the
valence  quark and proton  spin-dependent
structure functions $g_1(x,Q^2)$
is estimated within the instanton liquid model for QCD vacuum.
It is shown, that this
interaction leads
to a rather large violation of the Ellis-Jaffe sum rule.

\end{abstract}

\newpage

\section{Introduction}

One of the solution of the famous ''spin crisis''
(see recent review \ct{rev}),  is an assumption of the large
positive gluon polarization inside a nucleon  \ct{AA}.

Indeed,  one of the NLO analyses
of the polarized DIS world data on $g_1(x,Q^2)$, which was performed  to
extract the polarized parton densities in a nucleon,
has shown some  indication for
  a positive value of the gluon polarization \ct{FortRud}.
However, another NLO fit leads to the conclusion that
 this result may be sensitive to the  input shapes of the
polarized parton distribution at low normalization point \ct{Reya}.

The different  calculations of the gluon polarization inside
nucleon have been performed (see \ct{mank}) and
the contradiction between different approaches has
been found, not only  in the absolute value of the gluon polarization
but even in its sign.
For example it has been shown recently \ct{pol} that  the
instanton model for nonperturbative effects in QCD \ct{a5},\ct{a55}
rules out the positive value of the gluon polarization.
Therefore, it is possible that the gluon solution of
the "spin crisis"  is not the case.

Another way to resolve the proton spin problem is to take into
account the quark depolarization induced by nonperturbative
vacuum fluctuations of the gluon fields, so-called instantons
\ct{DorKoch}, \ct{Forte}.
The instantons describe the subbarrier transitions between different
classical QCD vacuum states, that  have different values of
quark helicities.
Therefore,  account of quark interaction with
instantons gives the direct way to obtain the magnitude of the
quark helicity non-conservation in QCD.

In this article we estimate the contribution from
 quark-quark \ct{Hooft} and quark-gluon interaction
\ct{koch1} induced by instantons
 to  the valence quark and proton $g_1(x,Q^2)$ structure functions.

\section{Quark-Quark and Quark-Gluon Interaction induced by
Instantons}

The instanton model
for QCD vacuum is widely used now in the description
of the nonperturbative effects in strong interaction
(see reviews \ct{a5}, \ct{a55}).
The existence of  instantons leads to a  nonperturbative
quark-quark  and quark-gluon interaction through the QCD vacuum,
which
has the following form \ct{Hooft},\ct{CDG}:
\ba
{\cal L}_{eff}&=&\int\prod_q(m_q\rho-2\pi^2\rho^3\bar q_R(1+\frac{i}{4}
\tau^aU_{aa^\prime}\bar\eta_{a^\prime\mu\nu}\sigma_{\mu\nu})q_L)
\nn\\
&\cdot &exp^{-\frac{2\pi^2}{g}\rho^2U_{bb^{\prime}}
\bar\eta_{b^\prime\gamma\delta}
G^b_{\gamma\delta}}
\frac{d\rho}{\rho^5}d_0(\rho)d\hat{o}
+R\longleftrightarrow L,
\label{e1}
\ea
 where $\rho$ is the instanton size, $\tau^a$ are the matrices of the
 $SU(2)_c$ subgroup of the $SU(3)_c$ colour group,
 $d_0(\rho)$ is the density of the instantons, $d\hat{o}$ stands
 for integration over the instanton orientation in colour space,
$\int d\hat{o}=1$,
$U$ is the orientation matrix of the instanton,
 $\bar\eta_{a\mu\nu}$ is the numerical t'Hooft symbol and
 $\sigma_{\mu\nu}=[\gamma_\mu,\gamma_\nu]/2$.

From Eq.\re{e1} one can obtain an effective quark-quark t'Hooft
interaction \ct{Hooft} which in the framework of instanton
liquid model reads  \ct{a5},\ct{koch1}
\begin{eqnarray}
{\cal L}_{eff}^{(2)}&=&
 F(k_1^2,k_2^2,k_3^2,k_4^2,\rho_c)\frac{4\pi^2\rho_c^2}{3}\sum_{i\neq j}
(\bar q_{iR}(k_1)q_{iL}(k_2)\bar q_{jR}(k_3)q_{jL}(k_4)\nn\\
&\cdot & (1+\frac{3}{8}(1-\frac{3}{4}\sigma_{\mu\nu}^i\sigma_{\mu\nu}^j)
t_u^at_d^a+(R\longleftrightarrow L))),
\la{e44}
\end{eqnarray}
where $F(k_1^2,k_2^2,k_3^2,k_4^2,\rho_c) $ is the form factor, which is
related to the Fourier transformation of the quark zero modes in
instanton field, $\rho_c\approx 1.6 GeV^2$ is an average instanton
size in vacuum, $i,j=u,d,s$.

Recently, it was shown that from Eq.\re{e1}
 a new type of the nonperturbative  quark-gluon  interaction can be
 obtained. This interaction has the form of
 anomalous chromomagnetic
quark-gluon interaction \ct{koch1}
\be
\Delta {\cal L_A}=
-i\mu_a
\sum_q\frac{g}{2m_q^*}\bar q\sigma_{\mu\nu}
t^a qG_{\mu\nu}^a,
\label{e4}
\ee
where
$m_q^*=2\pi^2\rho_c^2<0\mid \bar qq\mid 0>/3$ is a quark mass in
the instanton vacuum.

The value of the quark anomalous chromomagnetic moment
in the liquid instanton model is
\be
\mu_a=-\frac{f\pi}{2\alpha_s},
\label{a6}
\ee
 where $f=n_c\pi^2\rho_c^4$ is the so-called packing fraction of instantons
 in  vacuum.
The value of $n_c$ is connected with the value of the
gluon condensate by the formula
\be
n_c=<0\mid \alpha_sG_{\mu\nu}^a G_{\mu\nu}^a\mid 0>/16\pi
\approx 7.5~10^{-4}{\  }GeV^4.
\nn
\ee
The following estimate for
 the value of the anomalous quark chromomagnetic
moment has been obtained for $\rho_c=1.6GeV^{-1}$ in \ct{koch1}
\begin{equation}
\mu_a=-0.2.
\nonumber
\end{equation}

The principal difference between the instanton induced interaction \re{e1},
\re{e44} and the perturbative quark-gluon vertex
is the large quark helicity flip at instanton vertex, namely
$\Delta\Sigma=-2N_f$.  Therefore, this interaction  can be
responsible for a rather strong violation of
 Ellis-Jaffe sum rule \ct{EJ} for the first moment of the spin-dependent
structure function $g_1^p(x,Q^2)$.

\section{Instanton Contribution to  Valence Quark \\ and Proton Structure  Functions}

The graphs  which give the contribution
  to the   structure
functions from quark-quark interaction  \re{e44}
are presented in  Fig.1.

\begin{figure}[htb]
\centering
\epsfig{file=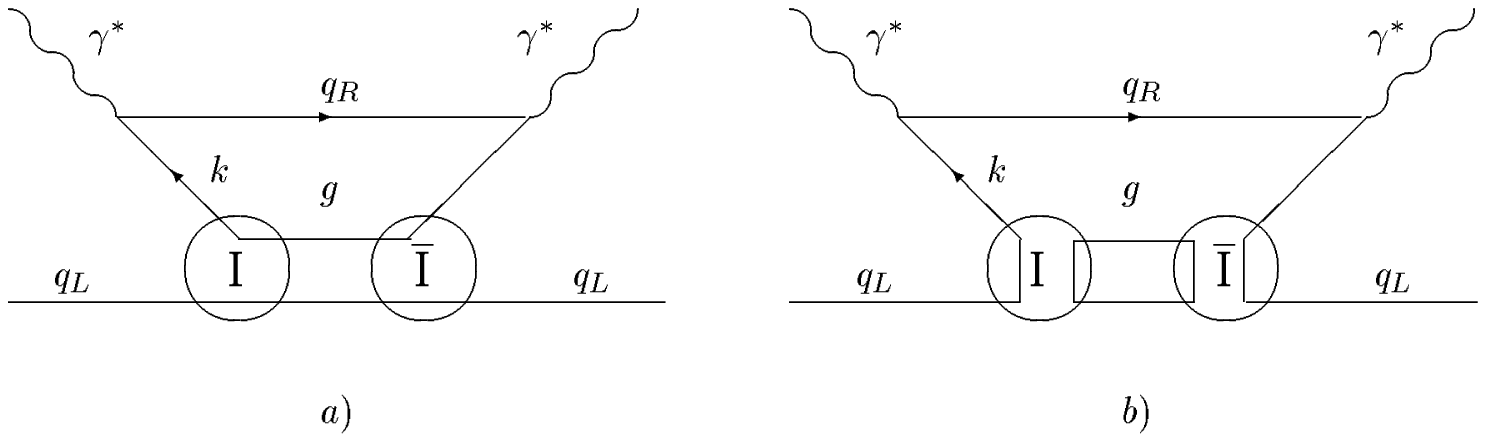,width=9cm}
\vskip 3cm
\caption{\it The instanton contribution to the  quark
distribution functions: $a)$ the contribution to the sea quark
distribution function; $b)$ the contribution to the valence quark
distribution function.
 The label $I(\bar I)$ denotes
instanton (antiinstanton).}

\end{figure}

 We calculate the contribution of this interaction
to the valence quark spin-dependent structure function
$g^q_1(x,Q^2)$ by using
the projection of the imaginary part of the forward Compton scattering
amplitude $T_{\mu\nu}$
\be
g^q_1(x,Q^2)=-\frac{ie_{\mu\nu\rho\sigma}p^\rho q^\sigma ImT_{\mu\nu}}{p.q},
\la{pro}
\ee
where $p$ is the momentum  of the valence quark in nucleon.
The straightforward  calculation of the contribution of the diagram given
by Fig.1b
\footnote{ The calculation has shown that the
contribution  of the diagram in Fig.1a is very small
and therefore can be neglected. The terms which are proportional
to the product of the colour matrix $t^a$ in \re{e44}
have $1/N_c$ suppression factor  and give the very
small contribution as well.}
leads to the result
\be
g_1^{q,q}(x,Q^2)=\frac{e_q^2\rho_c^4}{128}\int_0^{\frac{Q^2(1-x)}{4x}}dk_\bot^2
\\ \nn
\int_{\frac{-k_\bot^2}{1-x}}^{-\infty}dk^2\nn\\
\frac{(k_\bot^4+k_\bot^2k^2(1-x))}
{x^2k^4}F^2(\frac{|k|\rho_c}{2}),
\la{fig1b}
\ee
where
\be
F(z)=z\frac{d}{dz}[I_0(z)K_0(z)-I_1(z)K_1(z)],
\la{ff}
\ee
and $x=Q^2/2p.q$.

The graph which is responsible for the contribution from the quark-gluon
chromomagnetic interaction \re{e4} to structure functions is presented
in Fig.2.
\begin{figure}[htb]
\centering
\epsfig{file=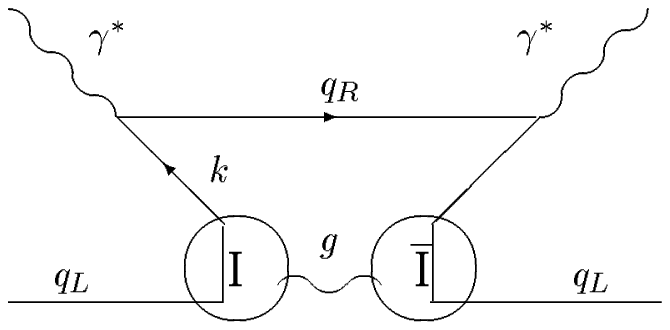,width=9cm}
\vskip 3cm
\caption{\it The instanton contributions to  proton
structure function from the anomalous
chromomagnetic interaction.}
\end{figure}
The calculation of the contribution of
diagram in Fig.2 leads to the result
\be
g_1^{q,g}(x,Q^2)=-\frac{e_q^2}{8}\frac{|\mu_a|\rho_c^2}{(1-x)}
\int_0^{\frac{Q^2(1-x)}{4x}}dk_\bot^2\\ \nn
\frac{F^2(k\rho_c/2)}{\sqrt{1-\frac{4xk_\bot^2}{(1-x)Q^2}}},
\la{fig1a}
\ee
where
$k^2=k_\bot^2/(1-x)$,
and the relation
$\alpha_s\mu_a^2/m_q^{*2}\rho_c^2=3\pi|\mu_a|/8$ \ct{pol}
has been used.

The  very interesting feature of the instanton contributions
\re{fig1b}  and \re{fig1a} is their specific $Q^2$ dependence.
At small $Q^2\ll 1/\rho_c^2$ they are proportional to $Q^2$  and for
$Q^2\gg 1/\rho_c^2$ they are constant.
 Therefore $Q^2$ dependence of the instanton
contribution to polarized  structure functions
{\it should be different} from $Log(Q^2/\Lambda^2)$ evolution of the
perturbative gluon corrections to DIS structure functions. The fundamental
reason for this feature is the quark spin-flip induced by instanton vertex,
which gives the extra powers of $k_\bot$ to the matrix element for forward
Compton scattering amplitude.
 As result, at $Q^2=0$ the instanton
contribution to $g_1(x,Q^2)$ is {\it zero}.

In the large $Q^2$ limit and small $x$ the contribution to $g_1^q(x)$
from quark-quark interaction (Fig.1b) has the anomalous $1/x^2$ behavior.
This behavior comes from the
point-like instanton vertex \re{e1}, \re{e44} which leads to
very fast growth of the $\bar q q$ cross section induced by instantons
with increasing the energy $S=(p-k)^2$. At high $S$ the contribution
of instantons should be small because the perturbative QCD
should work in this region. The imaginary part of the lower part
of the diagram Fig.1b is proportional to a sum of the imaginary
parts of the correlators of the pseudoscalar $j=\bar q\gamma_5q$
and scalar $j=\bar qq$ currents  ($s=\sqrt{S}$)
(see the  structure of Lagrangian \re{e44})
\be
\Pi(s)=i\int dxe^{isx}<0\|T\{j(x),j(0)\}\|0>,
\la{sr}
\ee
that have been analyzed in QCD sum rule approach, taking into account
direct instanton contribution \ct{shur},\ct{a5}.
Therefore, the threshold value for the energy
$S_0$ for cut-off of the instanton contribution  should approximately
equal to the value of duality interval in QCD sum rule for nonet of the
pseudoscalar and scalar mesons which is $S_0\approx 2 GeV^2 $ \ct{shur}.

The final formula for the  contribution from quark-quark
interaction  $g_1^q(x)$
in Bjorken limit $Q^2\rightarrow \infty$ can be written in
the following form
\be
g_1^q(x,Q^2)=\frac{e_q^2\rho_c^4}{128}\int_0^{S_0}dS\\ \nn
\int_{\frac{Sx}{1-x}}^{\infty}dk^2\nn\\
\frac{S(xS-k^2(1-x))}
{k^4}F^2(\frac{k\rho_c}{2}),
\la{fig}
\ee

Due to a cut-off in $S$, the $1/x$ divergence is absent now
and instanton contribution to first moment of $g_1(x)$ is finite.

In the same limit the contribution to $g_1^q(x)$, due to
quark-gluon chromomagnetic interaction, reads
\be
g_1^q(x)=-\frac{e_q^2}{4}|\mu_a|.
\la{ff1}
\ee

The sign of both contributions is {\it negative} and comes from the
{\it negative} quark polarization induced by instantons inside proton.

To calculate the contribution to proton structure function,
the simple convolution model for structure function has been used
\be
g_1^p(x)=\sum_q\int_x^1\frac{dy}{y}g_1^q(\frac{x}{y})\Delta q_V(y),
\nn
\ee
where $\Delta q_V(y)$ are the initial valence quark polarizations,
which were taken in the form
 \ba
\Delta u_V(x)&=&3.7(1-x)^3, {\ }\Delta d_V(x)=-1.3(1-x)^3,\label{val}
\ea
and  normalized
 to the experimental data
on the weak decay coupling constants of  hyperons
\be
g_A^3=\Delta u_V-\Delta d_V=1.25; {\ } g_A^8=\Delta u_V+\Delta d_V=0.6.
\label{coupl}
\ee

In Fig.3 the result of the  calculation of the  contribution
from quark-quark interaction induced by instanton to $g_1^p(x)$
in the region of Bjorken variable  $x>0.0001$
is presented.
\begin{figure}[htb]
\centering
\epsfig{file=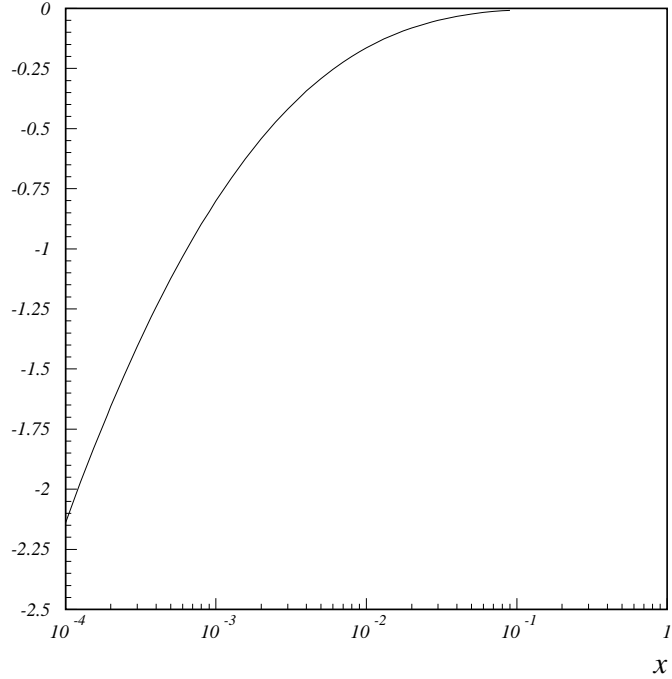,width=10cm}
\caption{\it  The   contribution to
 proton spin-dependent structure function
$g_1^p(x)$ due to quark-quark instanton induced interaction. }
\end{figure}

It should be mentioned that this contribution is negative and rather large,
especially in the low $x$ region.
The contribution
to the first moment of $g_1^p$  is
\be
\delta I^{p,q}_{inst}=\int_0^1dxg_1^{p,q}(x)=-0.007.
\la{first1}
\ee
This value is  approximately one fourth of the  observed violation of
the Ellis-Jaffe sum rule \ct{conf}.
 The remaining part of the violation
can be related to the contribution  from the anomalous quark-gluon
interaction induced by instantons.

In Fig.4  the result of the calculation of the contribution
to $g_1^p(x)$ structure function which comes from
 the quark-gluon chromomagnetic interaction induced by instantons
is shown.

\begin{figure}[htb]
\centering
\epsfig{file=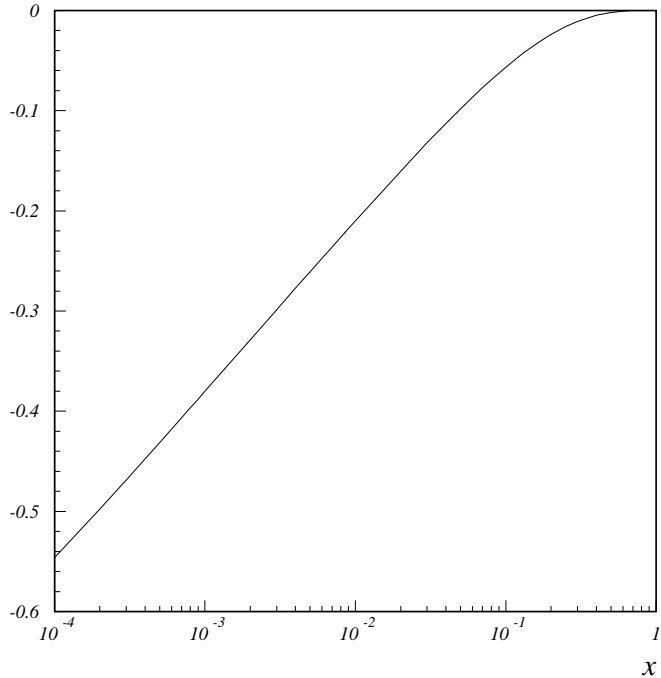,width=10cm}
\caption{\it  The   contribution to  the
 proton spin-dependent structure function
$g_1^p(x)$ due to the quark-gluon chromomagnetic interaction. }
\end{figure}

This contribution is also negative and has a more hard  $x$ dependence
than the contribution from the quark-quark interaction. Therefore
by taking into account the quark-gluon chromomagnetic interaction
induced by instantons we can explain the decrease of the
$g_1^p(x)$ structure function at large Bjorken variable $x$.

The contribution
to the first moment of $g_1^p$  from this interaction is
\be
\delta I^{p,g}_{inst}=\int_0^1dxg_1^{p,g}(x)=-0.019.
\la{first}
\ee

The total contribution from both interactions is
$\delta I^p_{inst}=-0.026$. This number one can compare with
the modern experimental data on the value of the violation
of the Ellis-Jaffe sum rule $\delta I^p_{exp}=-0.02\div-0.04 $
  \ct{conf}.

Thus, by taking into account the accuracy of the available
experimental data
and some ambiguities in the  extrapolations of the $g_1(x)$ to very low
$x$ region  in the current experiments, we can conclude that
the instanton model gives a rather good description of the
the Ellis-Jaffe sum rule violation
for proton. The prediction of the instanton model for the
neutron $g_1$ structure function
is very sensitive to the details of the violation of the
$SU(6)$ symmetry for the valence quark distribution function and
will be discussed elsewhere.

\section{Summary}

In summary,  the instanton induced  quark-quark and quark-gluon
interaction
leads  to a {\it large negative} contribution to the proton
spin-dependent structure
function $g_1^p(x,Q^2)$. This contribution allows us to explain
the observed violation of the Ellis-Jaffe sum rule.

\section*{Acknowledgements}

The author is
thankful to M.Anselmino, J.Bl\"umlein, A.E.Dorokhov, 
A.V.Efremov,R.L.Jaffe, E.Leader,
E.Reya, and T.Morii  for helpful
discussions.

This work was supported in part by the Heisenberg-Landau program and by the
Russian Foundation for Fundamental Research (RFFR) 96-02-18096.


\end{document}